\documentclass[aps,prl,twocolumn,showpacs,superscriptaddress]{revtex4-1} 
\usepackage{graphicx}
\usepackage{amsmath}
\usepackage{color}
\begin{document}

\title{Candidate quantum spin liquid in the Ce\textsuperscript{3+} pyrochlore stannate Ce$_2$Sn$_2$O$_7$}
\author{Romain Sibille}
\email[]{rom.sibille@gmail.com}
\affiliation{Laboratory for Developments and Methods, Paul Scherrer Institut, 5232 Villigen PSI, Switzerland}
\author{Elsa Lhotel}
\affiliation{Institut N\'eel, CNRS and Universit\'e Joseph Fourier, BP 166, 38042 Grenoble Cedex 9, France}
\author{Vladimir Pomjakushin}
\affiliation{Laboratory for Neutron Scattering and Imaging, Paul Scherrer Institut, 5232 Villigen PSI, Switzerland}
\author{Chris Baines}
\affiliation{Laboratory for Muon Spin Spectroscopy, Paul Scherrer Institut, 5232 Villigen PSI, Switzerland}
\author{Tom Fennell}
\email[]{tom.fennell@psi.ch}
\affiliation{Laboratory for Neutron Scattering and Imaging, Paul Scherrer Institut, 5232 Villigen PSI, Switzerland}
\author{Michel Kenzelmann}
\affiliation{Laboratory for Developments and Methods, Paul Scherrer Institut, 5232 Villigen PSI, Switzerland}

\begin{abstract}
We report the low temperature magnetic properties of Ce$_2$Sn$_2$O$_7$, a rare-earth pyrochlore.  
Our susceptibility and magnetization measurements show that due to the thermal isolation of a Kramers doublet ground state, 
Ce$_2$Sn$_2$O$_7$ has Ising-like magnetic moments of $\sim1.18$ $\mu_\mathrm{B}$.  The magnetic moments are confined to the local trigonal axes, as in a spin ice, but the exchange interactions are antiferromagnetic. 
Below 1 K the system enters a regime with antiferromagnetic correlations. In contrast to predictions for classical $\langle 111 \rangle$-Ising spins on the pyrochlore lattice, there is no sign of long-range ordering down to 0.02 K. Our results suggest that Ce$_2$Sn$_2$O$_7$ features an antiferromagnetic liquid ground state with strong quantum fluctuations.\end{abstract}

\maketitle

Quantum-mechanical phase coherence is a major theme of modern physics.  Various states with macroscopic quantum coherence such as superconductors~\cite{Josephson69}, superfluids ~\cite{Mayers2004}, fractional quantum Hall states ~\cite{Kellogg2004} and optically confined Bose-Einstein condensates ~\cite{Andrews1997} have been identified, all with remarkable macroscopic properties.  In insulators containing localized spin degrees of freedom, spin liquids can emerge~\cite{Balents:2010jx,Normand:2009wk}, which have no conventional order parameter associated with a broken symmetry, but whose defining character is a long-range entangled groundstate wavefunction~\cite{Isakov:2011fz,Wen:2002cy}.  Spin liquids are of great interest thanks to the remarkable collective phenomena that they can present, such as emergent gauge fields and fractional quasiparticle excitations~\cite{Gingras:2014ip,Moessner:2010ev}. Such states may also offer the possible application of coherent or topologically protected ground states in quantum information processing devices~\cite{Yao:2013ea}.

Quantum coherence of a spin system lacking symmetry-breaking order is well established in one-dimensional spin chains forming a spin fluid with a quantum coherence length almost an order of magnitude larger than the classical antiferromagnetic correlation length ~\cite{Xu2007}.  In higher dimensions two paradigms are employed, often simultaneously, to try to obtain a quantum spin liquid (QSL).  Firstly, for Heisenberg spins with S=1/2, where quantum mechanical corrections are most significant compared to classical states, quantum melting of the N\'eel ground state may be possible when spins pair into valence bond singlets~\cite{Lhuillier:BEVn7Mgc}.  The result may be a valence bond crystal (translationally ordered valence bonds) ~\cite{Korepin2010}, a resonating valence bond state (singlet configurations resonate around a plaquette) ~\cite{Anderson}, or a true spin liquid when valence bonds can be formed at all lengthscales so that the ground state wavefunction has a genuine long-range entanglement ~\cite{Liang1988, Balents:2010jx}.  Secondly, geometrically frustrated magnets are a natural landscape for liquid-like states of magnetic moments.  In two dimensions, the triangular and kagome lattices are important examples ~\cite{Lee2007,Yamashita2010,Pratt2011,Isono2014}, and neutron scattering experiments on the S=1/2 kagome lattice antiferromagnet ZnCu$_3$(OH)$_6$Cl$_2$ (herbertsmithite) have provided evidence of fractionalized excitations in a 2D QSL ~\cite{Han2012,DeVries:2009gh}.  In three dimensions QSLs are expected on the hyperkagome (e.g. in Na$_4$Ir$_3$O$_8$ ~\cite{Okamoto2007}) and pyrochlore lattices.  Despite the preponderance of S=1/2 spin liquid candidates mentioned above, recent work on pyrochlore spin liquid candidates such as Yb$_2$Ti$_2$O$_7$ ~\cite{Ross:2011tv,Chang:2012el},  Pr$_2$Zr$_2$O$_7$ ~\cite{Kimura:2013gj} and Pr$_2$Sn$_2$O$_7$ ~\cite{Zhou:2008cz} have illustrated how quantum effects can become important in materials where they may not be expected, i.e. in rare earth materials where crystal field effects lead to highly anisotropic magnetic moments.

The spin system of a pyrochlore with a thermally isolated doublet ground state can be described by a generalized Hamiltonian for effective $S=1/2$ spins \cite{Curnoe:2008gy,Ross:2011tv}. This Hamiltonian includes all symmetry-allowed near neighbour magnetic exchange interactions, with a leading interaction which establishes a classical ground state if acting alone, and competing transverse exchange terms that introduce quantum fluctuations.  Notably for Kramers ions, there is no requirement for these competing exchange terms to be small with respect to the leading term ~\cite{Gingras:2014ip}. A leading ferromagnetic interaction leads to a classical spin liquid ground state, the spin-ice state.  Exotic quantum phases are obtained as a function of the transverse terms ~\cite{Hermele:2004gg,Onoda:2010jf,Benton:2012ep,Savary:2012cq,Gingras:2014ip}:  
the quantum spin ice or $U(1)$ spin liquid, a disordered phase whose emergent properties are those of a $U(1)$-gauge theory~\cite{Hermele:2004gg,Benton:2012ep,Gingras:2014ip}, and the Coulombic ferromagnet, an ordered phase with deconfined spinons, whose existence is under debate \cite{Gingras:2014qsi}.

In rare earth pyrochlores with antiferromagnetic interactions, where the Ising magnetic moment points ``in'' or ``out'' of the tetrahedron (i.e. along the local $\langle111\rangle$ axis), the classical ground state is the ``all-in/all-out'' FeF$_3$ structure~\cite{Bramwell:1998wka, FeFe3}.  The introduction of strong quantum effects may melt the classical order to produce a  type of spin liquid, rather as in other unfrustrated quantum antiferromagnets.
In this Letter we report on Ce$_2$Sn$_2$O$_7$, a pyrochlore magnet based on Ce$^{3+}$ ($4f^1$, $^2F_{5/2}$). The local moments have $\langle111\rangle$ Ising anisotropy, and we find that although antiferromagnetic spin correlations develop below approximately 1 K, there is no sign of magnetic order down to 0.02 K.
The magnetic moment is small, suggesting that the magnetic dipolar couplings are much smaller than magnetic exchange interactions. This makes Ce$_2$Sn$_2$O$_7$ an excellent model material to look for novel exchange-induced QSLs on the pyrochlore lattice.

\begin{figure}
\centerline{\includegraphics[width=6.2cm]{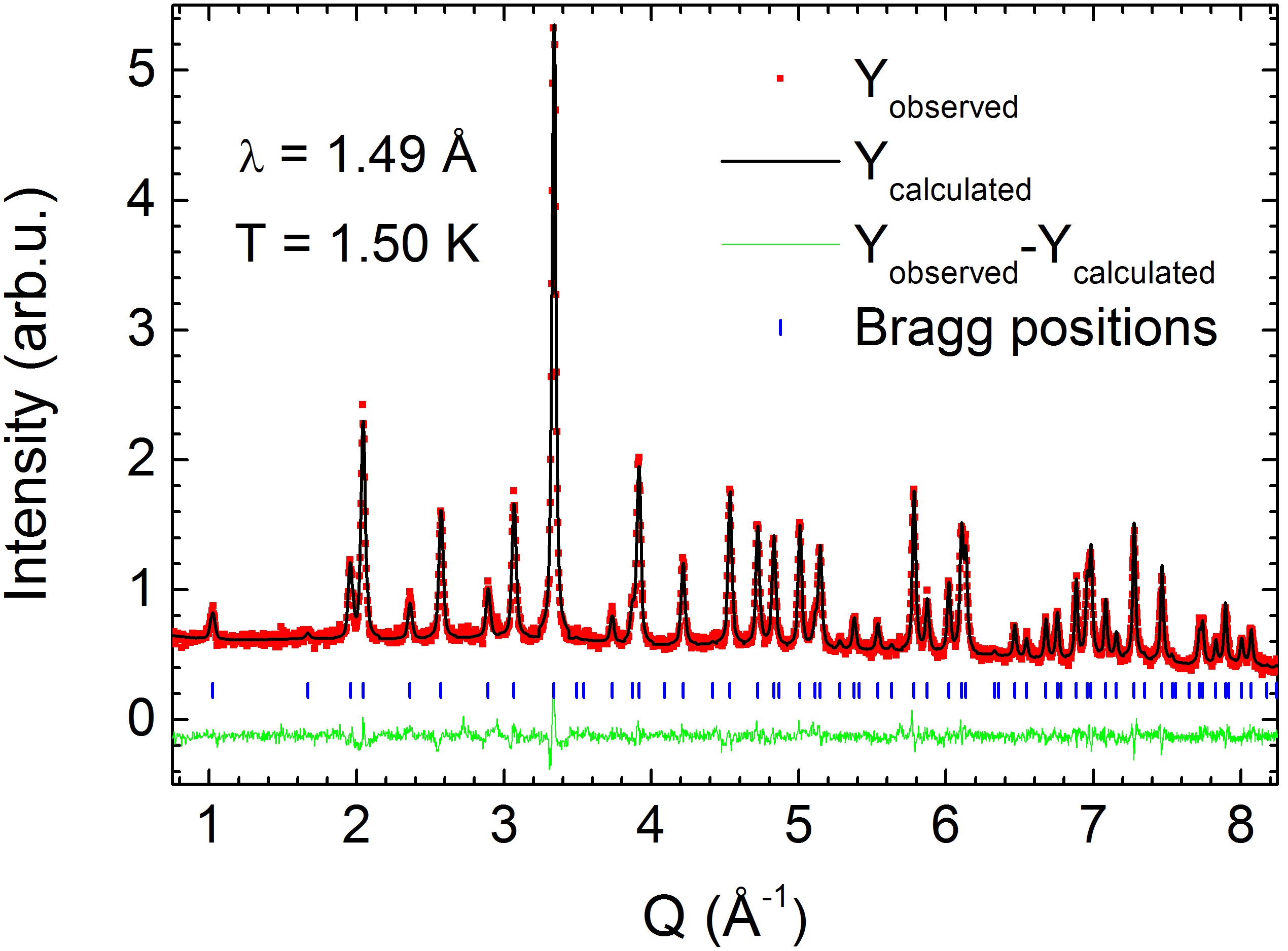}}
\caption{(color online) Rietveld refinement of neutron powder diffraction data (HRPT) collected at 1.5 K using an incident wavelength of 1.49 \AA. Fitted isotropic displacement parameters: $B_{\mathrm{Ce}}=0.87(4)$ \AA$^2$; $B_{\mathrm{Sn}}=0.79(3)$ \AA$^2$; $B_{\mathrm{O(48}f)}=1.08(2)$ \AA$^2$; $B_{\mathrm{O'(8}b)}=0.87(5)$ \AA$^2$. Conventional Rietveld factors (\%): $R_P=4.10$; $R_{WP}=5.19$; $R_{Bragg}=5.52$; $R_F=4.25$.}
\label{Fig.1}
\end{figure}

The low-temperature magnetic properties of Ce$^{3+}$ pyrochlores have been little studied, probably because of the difficulty to stabilize the magnetic Ce$^{3+}$ oxidation state in preference to the non-magnetic Ce$^{4+}$ ($4f^0$).
In Ce$_2$Sn$_2$O$_7$, a compound previously investigated for its oxygen storage capabilities~\cite{TOLLA:1999tq}, the trivalent rare-earth can be readily stabilized by taking advantage of a solid state oxydo-reductive reaction during which Sn$^0$ is oxidized to Sn$^{4+}$ while reducing Ce$^{4+}$ to the required Ce$^{3+}$. Our samples were produced using this method. Their oxygen stoichiometry, obtained from the thermogravimetric analysis procedure reported in Ref. \onlinecite{TOLLA:1999tq}, is $7.00\pm0.01$. The absence of excess oxygen indicates that all cerium cations are in their trivalent oxidation state and that diffraction data can be fitted assuming a stoichiometric formula unit. 
The Rietveld refinement of a neutron powder diffraction pattern is shown in Fig.~\ref{Fig.1} and gives the lattice parameter $10.6453(3)$ \AA~ at 1.5 K (space group: $Fd\bar{3}m$). The value of the atomic coordinate $\it{x}$ for the oxygen atom O(48$\it{f}$) is 0.3315(3), in the range of the typical values for A$_2$B$_2$O$_7$ compounds \cite{Gardner:2010fu}. The Ce-O(48$f$) bond length is $2.600\pm0.003$ \AA, close to the sum of the ionic radii ($2.68$ \AA), while Ce-O'(8$b$) bond (pointing along the local $\langle111\rangle$ direction) has a length of $2.305\pm0.003$ \AA, which is markedly shorter than $2.68$ \AA, as usually observed in rare-earth pyrochlores.  Attempts to refine antisite cation disorder ($0.5\pm2.5\%$) and oxygen Frenkel disorder ($0.36\pm0.16\%$), which can induce stuffing effects and disordered exchange interactions, respectively, did not provide evidence for structural defects.

Magnetization ($M$) data were measured in the temperature ($T$) range from 1.8 to 375 K in an applied magnetic field ($H$) of 100 Oe using a Quantum Design MPMS-XL SQUID magnetometer.  Additional magnetization, and $ac$-susceptibility, measurements were made as a function of temperature and field, from $T=0.07$ to $4.2$ K and from $H=0$ to $8\times10^{4}$ Oe, using SQUID magnetometers equipped with a miniature dilution refrigerator developed at the Institut N\'eel-CNRS Grenoble.  
The heat capacity ($C_p$) of a pelletized sample was measured down to 0.3 K using a Quantum Design PPMS.
Muon spin relaxation ($\mu$SR) measurements were performed at the LTF spectrometer
of the Swiss Muon Source, in the range from $T=0.02$ to $0.8$ K. Muons were longitudinally polarized and spectra were recorded in zero-field with earth-field compensation or in applied fields parallel to the beam.

\begin{figure*}
\centerline{\includegraphics[scale=0.224]{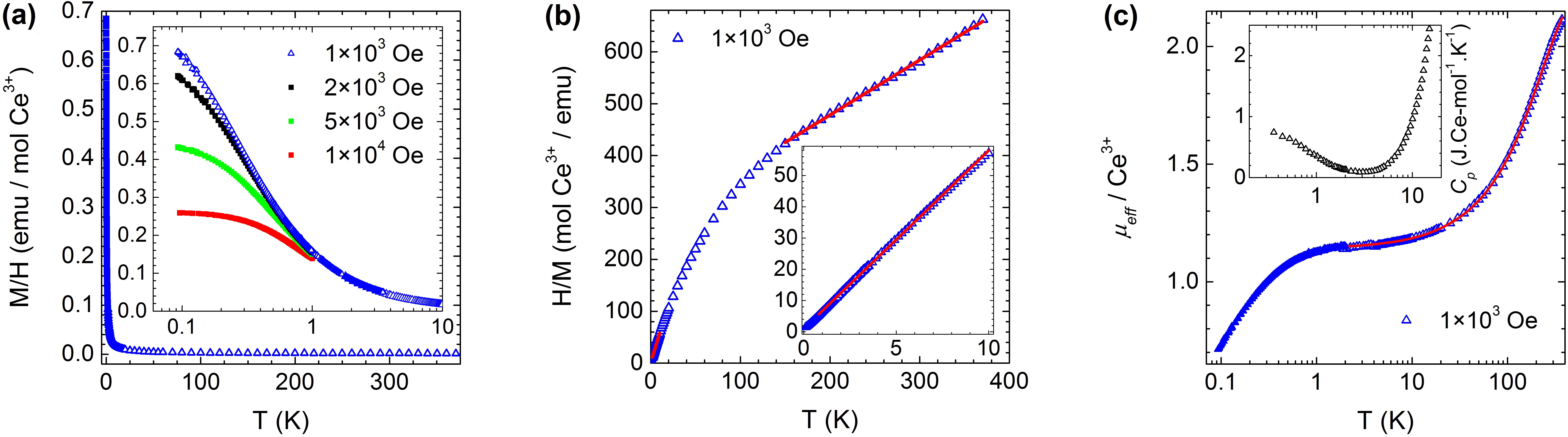}}
\caption{(color online) A: Magnetization $M$ as a function of temperature $T$ in a magnetic field $H= 1000$ Oe, plotted as the susceptibility $\chi(T)\sim M(T)/H$. The inset shows $M/H$ at several applied fields.  B: Inverse susceptibility $\chi^{-1}$ vs $T$. It exhibits two Curie-Weiss regimes (red lines) at high ($T>130$ K) and moderate (1 K $< T <$ 10 K) temperatures, and, in between, a regime which shows a curvature due to crystal field effects.  The inset shows a zoom on the moderate temperature Curie-Weiss regime; open and full symbols refer to data points from the high- and low-temperature magnetometers, respectively. C: Effective moment $\mu_{\mathrm{eff}}=((3k_{B}/N_{A}\mu^{2}_{B})\times\chi T)^{1/2}\sim 2.828\sqrt{\chi T}$ vs $T$. The red line is the fit, above 2 K, to the crystal field Hamiltonian. The inset in C shows the heat capacity on the same temperature scale as for the main panel.}
\label{Fig.2}
\end{figure*}

The magnetization divided by the applied field $M/H$, which is equal to susceptibility $\chi$ in the linear field regime, is shown as a function of the temperature $T$ over the full temperature range in Fig.~\ref{Fig.2}A.  The susceptibility increases continuously with decreasing temperature, and there is no evidence of any ordering transition (inset of Fig.~\ref{Fig.2}A). 
At high temperature, $T>130$ K, the inverse susceptibility  $\chi^{-1}$ (shown in Fig.~\ref{Fig.2}B) is almost linear, and a fit to the Curie-Weiss law yields a magnetic moment $\mu=2.75\pm0.20$ $\mu_\mathrm{B}/\mathrm{Ce}$, in reasonable agreement with the expected free ion value of 2.54 $\mu_\mathrm{B}/\mathrm{Ce}$, and $\theta_{CW}=-250\pm10$ K.  
This is an extremely large value for such a rare-earth material, where magnetic interactions are expected to be in the Kelvin range.  The large value of $\theta_{CW}$ can be attributed to crystal field effects, as shown by the strong curvature of $\chi^{-1}(T)$ below 100 K, indicating a change in the population of crystal field levels of the Ce$^{3+}$ ion.  At moderate temperatures, 1 K $< T <$ 10 K, a linear behavior is observed, and the Curie-Weiss fit to this part of $\chi^{-1}(T)$ gives a magnetic moment of $\mu=1.18\pm0.02$ $\mu_\mathrm{B}/\mathrm{Ce}$, which corresponds to the moment of the ground state doublet, and $\theta_{CW}=-0.25\pm0.08$ K. 
Fig.~\ref{Fig.2}C shows that the effective magnetic moment ($\mu_{\mathrm{eff}}=((3k_{B}/N_{A}\mu^{2}_{B})\times\chi T)^{1/2}\sim 2.828\sqrt{\chi T}$) approaches the free ion value at 375 K, and it falls to an approximate plateau of $1.18 ~\mu_\mathrm{B}$ in the range from $T=1$ to $10$ K.  At low temperature, $T < 1$ K, the effective moment drops.

The magnetic susceptibility was used to estimate the crystal field scheme. In the $LS$ coupling scheme, a crystal electric field with the $D_{3d}$ symmetry of the Ce$^{3+}$ site splits the $^2$F$_{5/2}$ free ion ground state into three Kramers doublets.  
However, the ground state multiplet $^2$F$_{5/2}$ alone does not allow to reproduce our experimental data. 
Instead, we used matrix elements of the crystal field Hamiltonian which are calculated by applying operators on the full basis of $f$-electron microstates (14 microstates in the case of Ce$^{3+}$)~\cite{Condon:1970}. This was accomplished using the computer program $Condon$ which takes into account the effect of the applied magnetic field and allows fitting the Wybourne coefficients of the ligand-field Hamiltonian on $\chi(T)$ data \cite{Schilder:2004vb}.  The refinement of six ligand-field parameters for the case of a $4f^1$ ion in $D_{3d}$ local symmetry to the susceptibility was realized between $T=1.8$ and $370$ K, and the resulting calculation of the single ion magnetic moment is shown in Fig.~\ref{Fig.2}C. The wavefunctions of the ground state Kramers doublet correspond to a linear combination of  $m_J=\pm3/2$ states.  The fitted coefficients result in energy levels at $50\pm15$ meV and $75\pm15$ meV, and four more levels distributed around 300 meV, which are all Kramers doublets. Although all levels consist of mixed ground ($^2F_{5/2}$) and first excited ($^2F_{7/2}$) multiplets, the lower levels are dominated by the $^2F_{5/2}$ term, while the upper four levels are dominated by the $^2F_{7/2}$ term. The local anisotropy axis of the crystal field levels yields a strong Ising anisotropy along the $\langle111\rangle$ axis. The intermultiplet splitting of $\approx300$ meV is a typical value for the transition between the ground and first excited multiplet in Ce$^{3+}$ compounds ~\cite{liu}.  This crystal field scheme of Ce$_2$Sn$_2$O$_7$ is generally consistent with the 
calculations using the $multiX$ computer program~\cite{Uldry:2012td} if the semi-empirical values for the spin-orbit coupling and crystal field scalers are adjusted accordingly.

\begin{figure}
\includegraphics[scale=0.202]{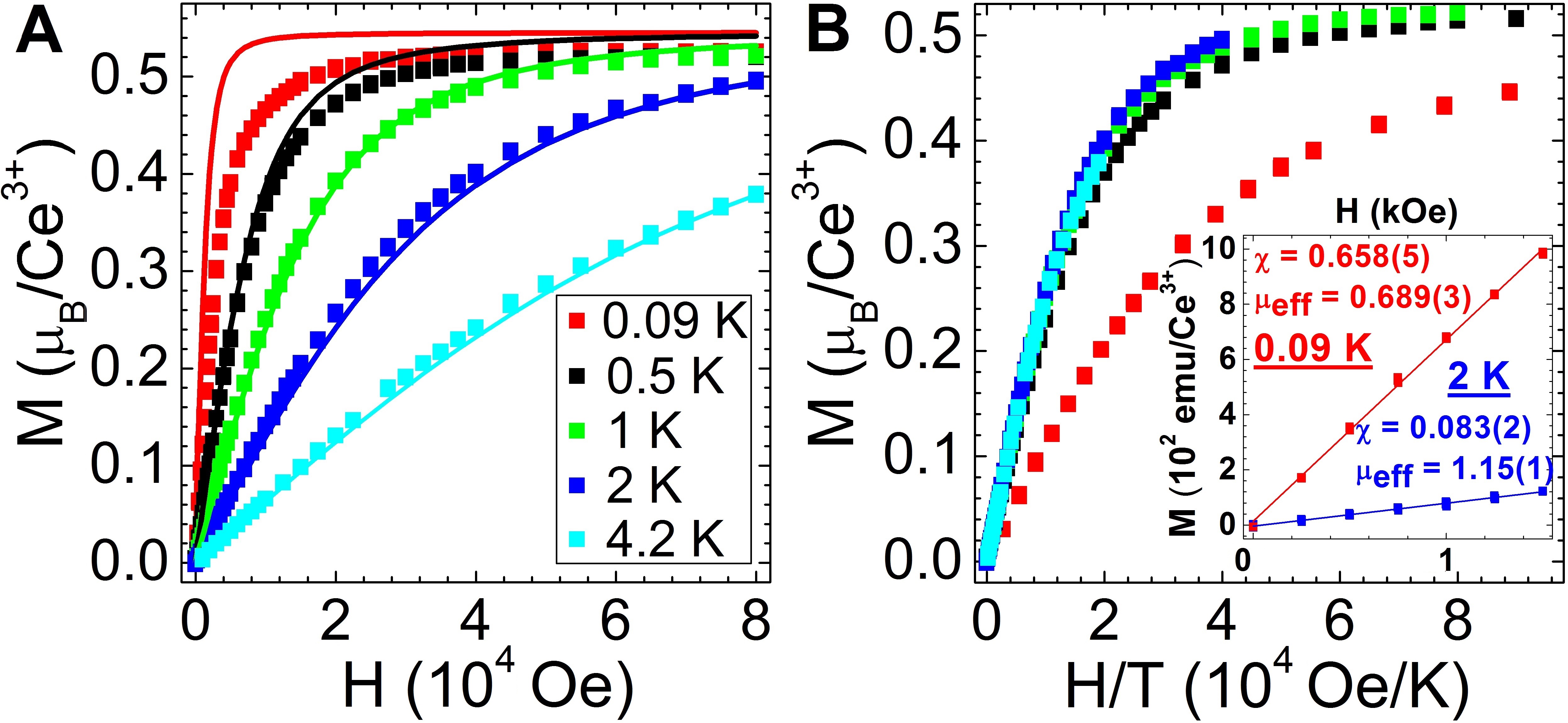}
\caption{(color online) Magnetization ($M$) recorded as a function of magnetic field ($H$). A: data in the form $M(H)$; lines are calculations for effective $S_{\mathrm{eff}}=1/2$ spins with $\langle111\rangle$ easy axis anisotropy and parameterized $g$-factor~\cite{Bramwell:2000tc}.  
B: data in the form $M(H/T)$, so that they collapse in the uncorrelated regime. The inset shows linear fits to low-field $M(H)$ data, enabling comparison to the effective moments on Fig.~\ref{Fig.2}C.
}
\label{Fig.3}
\end{figure}

In Fig.~\ref{Fig.3}A we show isothermal magnetization curves, $M(H)$, evidencing another striking feature. At moderate and low temperatures, i.e. in the plateau region of the effective moment and below, $M$ saturates at roughly half of the value of the effective magnetic moment observed in the moderate temperature plateau. This is reminiscent of the spin-ices Ho$_2$Ti$_2$O$_7$ and Dy$_2$Ti$_2$O$_7$ where, due to the important non-collinear local anisotropy, the low-temperature magnetization curves display a similar behavior ~\cite{Bramwell:2000tc}.  
Using a simple expression to model the magnetization of non-interacting Ising spins with local $\langle111\rangle$ easy-axis anisotropy and $S_{\mathrm{eff}}=1/2$ spins ~\cite{Bramwell:2000tc} our $M(H)$ data are satisfactorily reproduced for temperatures down to 1 K with a parameterized $g$-factor of 2.18.
The saturation of the magnetization up to applied fields as large as 8 T indicates a strong local anisotropy, as expected from the large energy gap to the first excited doublet.

We now examine the interactions amongst the Ce$^{3+}$ moments. The constant extracted from the Curie-Weiss fit at moderate temperatures ($\theta_{CW}=-0.25\pm0.08$ K) suggests antiferromagnetic interactions, but no ordering is observed in the magnetization data down to 0.07 K, as shown in the inset of Fig.~\ref{Fig.2}A. Evidence for antiferromagnetic correlations is provided by the value of the effective moment, which, below 1 K, falls below that of the ground state doublet, as shown in Fig.~\ref{Fig.2}C. Simultaneously, below this temperature, the isothermal $M(H)$ curves shown in Fig.~\ref{Fig.3}A depart from the single-ion form which reproduces well the curves at higher temperatures, thus confirming the onset of interactions and correlations. Moreover, the same magnetization curves are plotted as a function of $H/T$ in Fig.~\ref{Fig.3}B. Above 1 K, the curves collapse onto one another, as expected for uncorrelated spins ($T > \theta_{CW}$). Below 1 K, the curves increasingly deviate from this scaling, and their suppression for $T<1$ K relative to those for $T>1$ K supports the development of antiferromagnetic interactions. Finally, the drop of the effective magnetic moment at low temperature is corroborated by the linear fits to the low-field part of the $M(H)$ data (see inset of Fig.~\ref{Fig.3}B).

Heat capacity data, $C_p(T)$ (see inset of Fig.~\ref{Fig.2}C), show that the decrease of the effective magnetic moment at low temperature is accompanied by a rise in the heat capacity, which is usual when entering a short-range correlated state (see, e.g., the quantum spin ice candidates Pr$_2$Zr$_2$O$_7$ ~\cite{Kimura:2013gj} and Yb$_2$Ti$_2$O$_7$ ~\cite{Hayre:2013}).
This corroborates our claim that the variations observed in the magnetization data at sub-Kelvin temperatures are associated with cooperative phenomena. 
Given the energy levels derived from the fits to the susceptibility, it is unlikely that a Schottky anomaly is at the origin of the low temperature upturn in $C_p(T)$ (as was proposed for Ce$_2$Zr$_2$O$_7$ ~\cite{Popa:2008}), nor is any nuclear contribution expected for Ce$_2$Sn$_2$O$_7$ since all isotopes of cerium have nuclear spin of zero.

\begin{figure}
\includegraphics[scale=0.185]{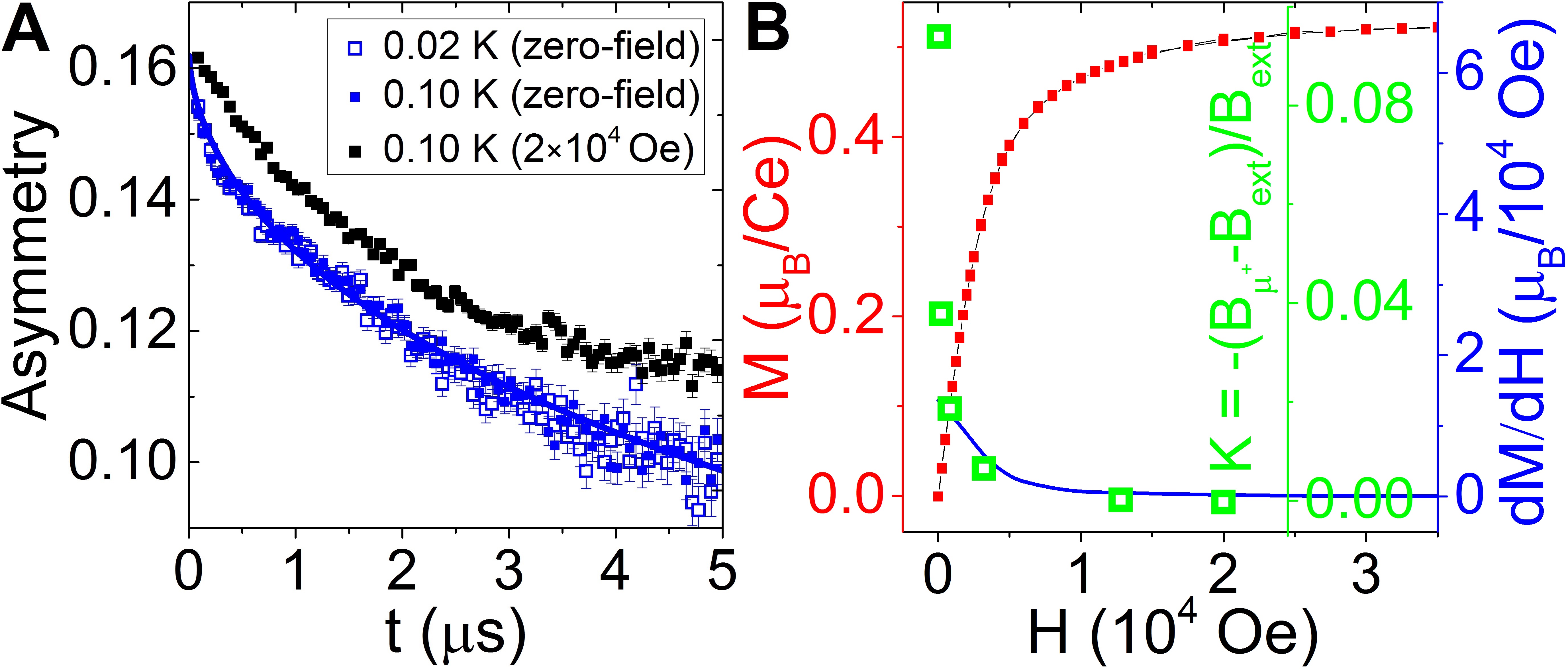}
\caption{(color online) Muon spin relaxation ($\mu$SR) experiments. A: zero-field and longitudinal field data in the usual histogram form $A(t)$; the blue line is a fit to the function $A(t)=A_0$$\times$$exp(-(t/T_1)^{\beta})+A_{bg}$. B: frequency shift ($K$) between the external field ($B_{ext}$) and the local field at the muon site ($B_{\mu^{+}}$) plotted at 0.1 K as a function of the external transverse field and scaled to the derivative of $M(H)$ at 0.09 K.}
\label{Fig.4}
\end{figure}

The $ac$-susceptibility (not shown) follows the $M(T)/H$ curve, has no frequency dependence down to 0.07 K in the range 1.11 - 211 Hz, and the out-of-phase signal remains unobservable in our experiments. The absence of any signature of magnetic freezing means, in the absence of long-range order, that the magnetic fluctuations are faster than the correlation times probed by the technique. $\mu$SR measurements were made in order to extend our study to lower temperatures and higher frequencies. Zero-field spectra were recorded at several temperatures between 0.02 K and 0.8 K. 
The zero-field data can be fitted with a stretched exponential relaxation, giving $1/T_1\sim0.05$ MHz and a stretched exponent $\beta\sim0.5$ (Fig.~\ref{Fig.4}A). No temperature dependence of the extracted $1/T_1$ and $\beta$ values was observed. 
The spin correlations are dynamic at low temperatures, because the relaxation function does not change in the presence of external longitudinal fields (see Fig.~\ref{Fig.4}A). In Fig.~\ref{Fig.4}B we show the frequency shift observed in transverse magnetic fields. It behaves like the magnetic susceptibility, indicating that the muon does not significantly perturb the system and that both techniques probe the same fluctuations, with differences at low-field which may be due to sensitive differences in internal field at the muon site and external field.

Classical spins on the pyrochlore lattice with $\langle111\rangle$ anisotropy together with competing Heisenberg exchange and dipolar interactions~\cite{denHertog:2000tc} lead to spin ices when the dipolar interaction ($D_{\mathrm{NN}}$) dominates, while a strong antiferromagnetic  exchange ($J_{\mathrm{eff}}$) stabilizes ``all-in/all-out'' long-range order. In Ce$_2$Sn$_2$O$_7$, using the moderate temperature effective moment, we can calculate $D_{\mathrm{NN}}=(5\mu_0\mu^2)/(12\pi r_{nn}^3)\sim0.025$ K, while the energy scale of the antiferromagnetic interactions, $|J_{\mathrm{eff}}|$, obtained from the whole set of magnetization measurements is about 0.5 K. Therefore, the resulting nearest neighbour interaction $J_{\mathrm{NN}}=|J_{\mathrm{eff}}|+D_{\mathrm{NN}}\sim |J_{\mathrm{eff}}|$, so that the system should be deep in the antiferromagnetic regime and a phase transition is expected at a temperature where the correlations become strong, around 0.5 K.
 Ce$_2$Sn$_2$O$_7$ does not conform to this prediction, suggesting that ``quantum fluctuations'' allow the system to evade the classicaly predicted magnetic order and to retain a correlated dynamical state down to a temperature at least one order of magnitude smaller than the temperature at which correlations are established. In Ce$_2$Sn$_2$O$_7$, the small value of the moment and its Kramers nature significantly enhance the quantum fluctuations on the pyrochlore lattice.

In summary, we suggest that Ce$_2$Sn$_2$O$_7$ is a model system to study a strongly correlated short-range antiferromagnetic state on the pyrochlore lattice. The magnetism features several important characteristics of exchange-based spin liquids - $\langle 111\rangle$ Ising spins coupled antiferromagnetically which become correlated but remain disordered to the lowest temperatures.  Ce$_2$Sn$_2$O$_7$ is structurally well ordered and based on an isolated Kramers doublet featuring a small moment.  This strongly Ising character of the localized magnetic moment is particularly important since all other pyrochlore materials with antiferromagnetic interactions featuring Kramers doublets either have $XY$ character (e.g Er$_2$Ti$_2$O$_7$ 
 ~\cite{Champion:2003ErTi,Zhitomirsky:2014:ErTi,Savary:2012:ErTi}) and/or larger moments (e.g. Nd$_2$Zr$_2$O$_7$ which appears to order below 300 mK ~\cite{blote}), while non-Kramers systems with magnetic doublets such as those based on Tb$^{3+}$ are further complicated by low-lying crystal fields {\cite{Gardner:2010fu}. 
Although classical calculations for the multi-axis Ising antiferromagnet on the pyrochlore lattice predict a conventional magnetic order, our data suggest that quantum fluctuations play an important role in destabilizing this ordering. It would be most interesting if theory could further our understanding of the Ising antiferromagnet on the pyrochlore lattice in the extreme quantum limit.

\acknowledgments{We thank C. Paulsen for the use of his magnetometers; P. Lachkar for help with the PPMS; B. Delley for help with $\it{multiX}$; H. Schilder for answering questions concerning {\it Condon}; and P. Holdsworth, P. A. McClarty and M. D. Le for informative discussions. We acknowledge funding from the European Community's Seventh Framework Program (grants 290605, COFUND: PSI-FELLOW; and 228464, Research Infrastructures under the FP7 Capacities Specific Programme, MICROKELVIN), and the Swiss NSF (grant 200021\_140862, Quantum Frustration in Model Magnets). Neutron scattering experiments were carried out at the continuous spallation neutron source SINQ at the Paul Scherrer Institut at Villigen PSI in Switzerland.}

\end{document}